\documentclass[10pt]{article}
\usepackage{latexsym,graphicx}
\usepackage{latexsym,graphicx}
\newcommand{\be}{\begin{equation}}
\newcommand{\ee}{\end{equation}}
\def\n{\noindent}
\catcode `\@=11 \catcode `\@=12
\begin{document}
\begin{center}
\large{LYRA'S COSMOLOGY OF INHOMOGENEOUS UNIVERSE WITH ELECTROMAGNETIC FIELD}\\
\vspace{6mm}
\normalsize{ANIL KUMAR YADAV} \\
\vspace{2mm}
\normalsize{\textit{Department of Physics, Anand Engineering College, Keetham, Agra-282 007, India \\ 
E-mail : abanilyadav@yahoo.co.in, anilyadav.physics@gmail.com}}\\
\end{center}
\vspace{6mm}

\n The plane-symmetric inhomogeneous cosmological models of
perfect fluid distribution with electro-magnetic field is obtained in the framework of Lyra's geometry.
To get the deterministic solution, I have considered $ A=f(x)\nu(t) $, $ B=g(x)\mu(t) $ and $ C=h(x)\mu(t) $,
where A, B and C are metric coefficients. It has been found that the solutions generalize the solution obtained
 by Pradhan, Yadav and Singh $ \left(2007\right) $ and are consistent
with the recent observations of type Ia supernovae. A detailed study of physical and kinematical properties of the 
models have been carried out.

\vspace{2mm}

\n Key words: Cosmology, Electromagnetic field, Inhomogeneous universe,
Lyra's geometry \\
\n PACS: 98.80.Jk, 98.80.-k \\
\begin{center}
\large{1. \textit{Introduction and motivations}}
\end{center}

In recent years, our knowledge of cosmology has improved remarkably by various experimental and theoretical results.
The universe is spherically symmetric and the matter distribution in it is on the whole isotropic and homogeneous.
But during the early stage of evolution, it is unlikely that it could have such a smoothed out picture so I consider
plane symmetry which provides an opprotunity for the study of inhomogegeneity.The present study deals with plane 
symmetric inhomogeneous models within the framework of Lyra's geometry in presence of electromagnetic field.
The essential difference between between the cosmological theories based on Lyra's geometry and Riemannian geometry 
lies in the fact that the constant vector displacement field $ \beta $ arises naturally from the concept of gauge in 
Lyra's geometry, where as the cosmological constant $ \Lambda $ was introduced in ad hoc fashion in the usual treatment.
Currently the study of gauge function and cosmological constant have gained renewed interest due to their application
in structure formation in the universe.
\newline
\par  
Einstein introduced his general theory of relativity in which gravitation is described in terms of geometry of 
space time. Einstein's idea of geometrizing gravitation in the form of general
theory of relativity inspired the idea of geometrizing other physical
fields. Shortly after Einstein's general theory of relativity Weyl
\cite{ref1} suggested the first so-called unified field which is a
geometrized theory of gravitation and electromagnetism. But this
theory was never taken seriously because it was based on the concept
of non-integrability of length transfer. Lyra \cite{ref2} proposed a
modification of Riemannian geometry by introducing a gauge function
which removes the non-integrability condition of the length of a
vector under parallel transport. In consecutive investigations Sen
\cite{ref3}, Sen and Dunn \cite{ref4} proposed a new scalar-tensor
theory of gravitation and constructed an analog of the Einstein
field equations based on Lyra's geometry. It is thus possible
\cite{ref3} to construct a geometrized theory of gravitation and
electromagnetism much along the lines of Weyl's ``unified'' field
theory without, however, the inconvenience of non-integrability
length transfer.
\newline
\par
Halford \cite{ref5} has pointed out that the constant vector
displacement field $\phi_i$ in Lyra's geometry plays the role of
cosmological constant $\Lambda$ in the normal general relativistic
treatment. It is shown by Halford \cite{ref6} that the scalar-tensor
treatment based on Lyra's geometry predicts the same effects, within
observational limits, as the Einstein's theory. Several authors Sen
and Vanstone \cite{ref7}, Bhamra \cite{ref8}, Karade and Borikar
\cite{ref9}, Kalyanshetti and Wagmode \cite{ref10}, Reddy and
Innaiah \cite{ref11}, Beesham \cite{ref12}, Reddy and Venkateswarlu
\cite{ref13}, Soleng \cite{ref14}, have studied cosmological models
based on Lyra's manifold with a constant displacement field vector.
However, this restriction of the displacement field to be constant
is merely one of convenience and there is no a priori reason for it.
Beesham \cite{ref15} considered FRW models with time dependent
displacement field. He has shown that by assuming the energy density
of the universe to be equal to its critical value, the models have
the $k=-1$ geometry. Singh and Singh \cite{ref16}$-$ \cite{ref19},
Singh and Desikan \cite{ref20} have studied Bianchi-type I, III,
Kantowaski-Sachs and a new class of cosmological models with time
dependent displacement field and have made a comparative study of
Robertson-Walker models with constant deceleration parameter in
Einstein's theory with cosmological term and in the cosmological
theory based on Lyra's geometry. Soleng \cite{ref14} has pointed out
that the cosmologies based on Lyra's manifold with constant gauge
vector $\phi$ will either include a creation  field and be equal to
Hoyle's creation field cosmology \cite{ref21}$-$ \cite{ref23} or
contain a special vacuum field which together with the gauge vector
term may be considered as a cosmological term. In the latter case
the solutions are equal to the general relativistic cosmologies with
a cosmological term.
\newline
\par
The occurrence of magnetic fields on galactic scale is
well-established fact today, and their importance for a variety of
astrophysical phenomena is generally acknowledged as pointed out by
Zeldovich et al. \cite{ref24}. Also Harrison \cite{ref25} has
suggested that magnetic field could have a cosmological origin. As a
natural consequences, we should include magnetic fields in the
energy-momentum tensor of the early universe. The choice of
anisotropic cosmological models in Einstein system of field
equations leads to the cosmological models more general than
Robertson-Walker model \cite{ref26}. 
Strong magnetic fields can be created due to adiabatic compression
in clusters of galaxies. Primordial asymmetry of particle (say
electron) over antiparticle (say positron) have been well
established as C P (charged parity) violation. Asseo and Sol
\cite{ref27} speculated the large-scale inter galactic magnetic
field and is of primordial origin at present measure $10^{-8}$ G and
gives rise to a density of order $10^{-35} g cm^{-3}$. The present
day magnitude of magnetic energy is very small in comparison with
the estimated matter density, it might not have been negligible
during early stage of evolution of the universe. FRW models are
approximately valid as present day magnetic field is very small. The
existence of a primordial magnetic field is limited to Bianchi Types
I, II, III, $VI_{0}$ and $VII_{0}$ as shown by Hughston and Jacobs
\cite{ref28}. Large-scale magnetic fields give rise to anisotropies
in the universe. The anisotropic pressure created by the magnetic
fields dominates the evolution of the shear anisotropy and it decays
slower than if the pressure was isotropic \cite{ref29,ref30}. Such
fields can be generated at the end of an inflationary epoch
\cite{ref31}$-$\cite{ref33}. Anisotropic magnetic field models have
significant contribution in the evolution of galaxies and stellar
objects. 
 Recently, Pradhan et al. \cite{ref34}, Casama et al. \cite{ref35},
Rahaman et al. \cite{ref36}, Bali and Chandani \cite{ref37}, Kumar
and Singh \cite{ref38}, Singh \cite{ref39} and Rao, Vinutha and
Santhi \cite{ref40} have studied cosmological models based on Lyra's
geometry in various contexts. Later on Rahaman et al. \cite{ref41,ref42}, Pradhan et al. \cite{ref43,ref44} obtained
some inhomogeneous cosmological models in Lyra's geometry. Motivated by these researches, in this
paper, I have studied plane-symmetric inhomogeneous cosmological
models in presence of magnetic field with the framework of Lyra's
geometry and also discussed the thermodynamical behaviour of universe. 
\begin{center}
\large{2. \textit{The Metric and field equations}}
\end{center}
We consider the plane-symmetric metric in the form
\begin{equation}
\label{eq1}
ds^{2} = A^{2}(dx^{2} - dt^{2}) + B^{2} dy^{2} + C^{2} dz^{2},
\end{equation}
where $A$, $B$ and $C$ are
functions of $x$ and $t$. The energy momentum tensor is taken as
\begin{equation}
\label{eq2}
T^{j}_{i} = (\rho + p)u_{i} u^{j} + p g^{j}_{i} +  E^{j}_{i},
\end{equation}
where $\rho$ and $p$ are, respectively, the energy density and
pressure of the cosmic fluid, and $u_{i}$ is the fluid four-velocity
vector satisfying the condition
\begin{equation}
\label{eq3}
u^{i} u_{i} = -1, ~ ~  u^{i} x_{i} = 0.
\end{equation}
In Eq. (\ref{eq2}), $E^{j}_{i}$ is the electromagnetic field given
by Lichnerowicz \cite{ref45}
\begin{equation}
\label{eq4}
E^{j}_{i} = \bar{\mu}\left[h_{l}h^{l}\left(u_{i}u^{j} + \frac{1}{2}g^{j}_{i}\right)
- h_{i}h^{j}\right],
\end{equation}
where $\bar{\mu}$ is the magnetic permeability and $h_{i}$ the magnetic flux vector
defined by
\begin{equation}
\label{eq5}
h_{i} = \frac{1}{\bar{\mu}} \, {^*}F_{ji} u^{j},
\end{equation}
where the dual electromagnetic field tensor $^{*}F_{ij}$ is defined
by Synge \cite{ref46}
\begin{equation}
\label{eq6}
^{*}F_{ij} = \frac{\sqrt{-g}}{2} \epsilon_{ijkl} F^{kl}.
\end{equation}
Here $F_{ij}$ is the electromagnetic field tensor and $\epsilon_{ijkl}$ is the
Levi-Civita tensor density.\\
The co-ordinates are considered to be comoving so that $u^{1}$ = $0$
= $u^{2}$ = $u^{3}$ and $u^{4} = \frac{1}{A}$. If we consider that
the current flows  along the $z$-axis, then $F_{12}$ is the only
non-vanishing component of $F_{ij}$. The Maxwell's equations
\begin{equation}
\label{eq7}
F_[ij;k] = 0,
\end{equation}
\begin{equation}
\label{eq8}
\left[\frac{1}{\bar{\mu}}F^{ij}\right]_{;j} = 4 \pi J^{i},
\end{equation}
require that $F_{12}$ is the function of x-alone. We assume that the magnetic
permeability is the functions of $x$ and $t$ both. Here the semicolon represents
a covariant differentiation. \\

The field equations, in normal gauge for Lyra's manifold, obtained by Sen
\cite{ref4} as
\begin{equation}
\label{eq9}
R_{ij} - \frac{1}{2} g_{ij} R + \frac{3}{2} \phi_i \phi_j
- \frac{3}{4} g_{ij} \phi_k \phi^k = - 8 \pi T_{ij},
\end{equation}
where $\phi_{i}$ is the displacement vector defined as
\begin{equation}
\label{eq10}
\phi_{i} = (0, 0, 0, \beta(t))
\end{equation}
and other symbols have their usual meaning as in Riemannian geometry. \\

For the line-element (\ref{eq1}), the field Eq. (\ref{eq9}) with
Eqs. (\ref{eq2}) and (\ref{eq10}) lead to the following system of
equations
\[
\frac{1}{A^{2}}\Biggl[- \frac{B_{44}}{B} - \frac{C_{44}}{C} +
\frac{A_{4}}{A} \left(\frac{B_{4}}{B} + \frac{C_{4}}{C}\right) +
\frac{A_{1}}{A}\left(\frac{B_{1}}{B} + \frac{C_{1}}{C}\right) - 
\]
\begin{equation}
\label{eq11}
\frac{B_{4}C_{4}}{BC} + \frac{B_{1}C_{1}}{BC}\Biggr] - \frac{3}{4}\beta^{2} 
= 8 \pi \left(p + \frac{F^{2}_{12}}{2\bar{\mu} A^{2} B^{2}} \right),
\end{equation}
\[
\frac{1}{A^{2}}\left[\frac{A_{11}}{A} + \frac{C_{11}}{C} - \frac{A_{44}}{A} 
- \frac{C_{44}}{C} - \left(\frac{A_{1}}{A}\right)^{2} + \left(\frac{A_{4}}
{A}\right)^{2}\right]
\]
\begin{equation}
\label{eq12}
 - \frac{3}{4}\beta^{2}   =  8 \pi \left(p +
\frac{F^{2}_{12}}{2\bar{\mu} A^{2} B^{2}} \right),
\end{equation}
\[
\frac{1}{A^{2}}\left[\frac{A_{11}}{A} + \frac{B_{11}}{B} - \frac{A_{44}}{A} 
- \frac{B_{44}}{B} - \left(\frac{A_{1}}{A}\right)^{2} + \left(\frac{A_{4}}
{A}\right)^{2}\right]
\]
\begin{equation}
\label{eq13}
 - \frac{3}{4}\beta^{2}   =  8 \pi \left(p +
\frac{F^{2}_{12}}{2\bar{\mu} A^{2} B^{2}} \right),
\end{equation}
\[
\frac{1}{A^{2}}\Biggl[- \frac{B_{11}}{B} - \frac{C_{11}}{C}
+ \frac{A_{1}}{A}\left(\frac{B_{1}}{B} + \frac{C_{1}}{C}\right)
+ \frac{A_{4}}{A} \left(\frac{B_{4}}{B} + \frac{C_{4}}{C}\right) 
- \frac{B_{1}C_{1}}{BC} + 
\]
\begin{equation}
\label{eq14}
 \frac{B_{4} C_{4}}{B C}\Biggr] + \frac{3}{4}\beta^{2} = 8 \pi \left(\rho + 
\frac{F^{2}_{12}}{2\bar{\mu} A^{2} B^{2}}\right),
\end{equation}
\begin{equation}
\label{eq15}
\frac{B_{14}}{B} + \frac{C_{14}}{C} - \frac{A_{1}}{A}\left(\frac{B_{4}}{B} 
+ \frac{C_{4}}{C}\right) - \frac{A_{4}}{A}\left(\frac{B_{1}}{B} +
\frac{C_{1}}{C}\right) = 0,
\end{equation}
where the sub indices $1$ and $4$ in A, B, C and elsewhere denote
ordinary differentiation with respect to $x$ and $t$ respectively.
\begin{center}
\large{3. \textit{Solutions and field equations}}
\end{center}
Equations (\ref{eq11}) - (\ref{eq13}) lead to
\[
\left(\frac{A_{4}}{A}\right)_{4} - \frac{B_{44}}{B} + \frac{A_{4}}{A}
\left(\frac{B_{4}}{B}
+ \frac{C_{4}}{C}\right) - \frac{B_{4}C_{4}}{BC} =
\]
\begin{equation}
\label{eq16} \left(\frac{A_{1}}{A}\right)_{1} + \frac{C_{11}}{C} -
\frac{A_{1}}{A} \left(\frac{B_{1}}{B} + \frac{C_{1}}{C}\right) -
\frac{B_{1}C_{1}}{BC} = \mbox{a (constant)}
\end{equation}
and
\begin{equation}
\label{eq17}
 \frac{8\pi F^{2}_{12}}{\bar{\mu}B^{2}} = \frac{B_{44}}{B} - 
\frac{B_{11}}{B} + \frac{C_{11}}{C} - \frac{C_{44}}{C}.
\end{equation}
Eqs. (\ref{eq10}) - (\ref{eq14}) represent a system of five 
equations in seven unknowns $A$, $B$, $C$, $\rho$, $p$, $\beta$  
and $F_{12}$. For the complete determination of these
unknowns one more condition is needed. As in the case of 
general-relativistic cosmologies, the introduction of 
inhomogeneities into the cosmological equations produces a 
considerable increase in mathematical difficulty: non-linear 
partial differential equations must now be solved. In practice, 
this means that we must proceed either by means of approximations 
which render the non-linearities tractable, or we must introduce 
particular symmetries into the metric of the space-time in order 
to reduce the number of degrees of freedom which the inhomogeneities 
can exploit. In the present case, we assume that the metric is Petrov 
type-II non-degenerate. This requires that
\[
\left(\frac{B_{11} + B_{44} + 2B_{14}}{B}\right) - 
\left(\frac{C_{11} + C_{44} + 2C_{14}}{C}\right) =
\]
\begin{equation}
\label{eq18}
\frac{2(A_{1} + A_{4})(B_{1} + B_{4})}{AB} - \frac{2(A_{1} +
A_{4})(C_{1} + C_{4})}{AC}.
\end{equation}
Let us consider that
\[
A = f(x)\nu(t),
\]
\[
B = g(x)\mu(t),
\]
\begin{equation}
\label{eq19}
C = h(x)\mu(t).
\end{equation}
Using (\ref{eq19}) in (\ref{eq14}) and (\ref{eq17}), we get
\begin{equation}
\label{eq20} \frac{\frac{g_{1}}{g} +
\frac{h_{1}}{h}}{\frac{f_{1}}{f}} = \frac{\frac{2\mu_{4}}
{\mu}}{\frac{\mu_{4}}{\mu}- \frac{\nu_{4}}{\nu}} = \mbox{b
(constant)}
\end{equation}
and
\begin{equation}
\label{eq21} \frac{\frac{g_{11}}{g} +
\frac{h_{11}}{h}}{\frac{g_{1}}{g} - \frac{h_{1}}{h}}
-\frac{2f_{1}}{f}=2\left(\frac{\mu_{4}}{\mu} -
\frac{\nu_{4}}{\nu}\right)  = \mbox{L (constant)}.
\end{equation}
Eq. (\ref{eq20}) leads to
\begin{equation}
\label{eq22}
f = n(gh)^{\frac{1}{b}}, \, \, \mbox{$b \ne 0$}
\end{equation}
and
\begin{equation}
\label{eq23}
\mu = m\nu^{\frac{b}{b - 2}},
\end{equation}
where $m$ and $n$ are constants of integration. \\
From Eqs. (\ref{eq15}), (\ref{eq18}) and (\ref{eq19}), we have
\begin{equation}
\label{eq24}
\frac{1}{b}\frac{g_{11}}{g} + \left(\frac{1 + b}{b}\right)
\frac{h_{11}}{h} - \frac{2}{b}\left(\frac{g^{2}_{1}}
{g^{2}} + \frac{h^{2}_{1}}{h^{2}}\right) - \frac{(2 + b)}{b}
\frac{g_{1}h_{1}}{gh} = a
\end{equation}
and
\begin{equation}
\label{eq25}
\frac{2}{b}\left(\frac{\mu_{44}}{\mu} + \frac{\mu^{2}_{4}}
{\mu^{2}}\right) = - a.
\end{equation}
Let us assume
\begin{equation}
\label{eq26}
g = e^{U + W}, ~ ~ ~ h = e^{U - W}.
\end{equation}
Eqs. (\ref{eq20}) and (\ref{eq25}) lead to
\begin{equation}
\label{eq27}
W_{1} = M\exp{\left(Lx + \frac{2(2 - b)}{b}U\right)},
\end{equation}
where $M$ is an integrating constant. From Eqs. (\ref{eq23}), 
(\ref{eq25}) and (\ref{eq26}),
we have
\[
\left(\frac{2 + b}{b}\right) U_{11} - \frac{4}{b}U^{2}_{1} -
2bM \exp{\left(Lx + \frac{2(2 - b)}{b}U\right)} -
\]
\begin{equation}
\label{eq28}
ML\exp{\left(Lx + \frac{2(2 - b)}{b}U\right)} + 2M^{2}\exp{\left(2Lx +
\frac{4(2 - b)}{b}U\right)} = a.
\end{equation}
Eq. (\ref{eq27}) leads to
\begin{equation}
\label{eq29}
U = \frac{Lbx}{2(b - 2)}, ~ ~ ~ b \ne 2
\end{equation}
Eq. (\ref{eq26}) and eq. (\ref{eq28}) lead to
\begin{equation}
\label{eq30}
W = M x + \log{N},
\end{equation}
where $N$ is the constant of integration. \\
Eq. (\ref{eq24}) leads to
\begin{equation}
\label{eq31}
\mu = \left[ \begin{array}{ll}
            \ell \cosh^{\frac{1}{2}}(\sqrt{\mid \alpha \mid} t + t_{0})
            & \mbox { when $ab<0$}\\
 \ell \cos^{\frac{1}{2}}(\sqrt{\alpha} t + t_{0})
            & \mbox { when $ab>0$}\\
            
(c_{1} t + t_{0})^{\frac{1}{2}}
 & \mbox { when $ab=0$}
  \end{array} \right.
\end{equation}
where $\alpha = ab$, $\ell$ is constant and $c_{1}$, $t_{0}$ are
constants of integration. Now we consider the following three cases.
\begin{center}
\large{4. \textit{Case(i): $ab < 0$}}
\end{center}
In this case we obtain
\begin{equation}
\label{eq32}
f = n \exp{\left(\frac{Lx}{(b - 2)}\right)},
\end{equation}
\begin{equation}
\label{eq33}
\mu = \ell \cosh^{\frac{1}{2}}(\sqrt{\mid \alpha \mid} t + t_{0}),
\end{equation}
\begin{equation}
\label{eq34}
\nu = r \cosh^{\frac{(b - 2)}{2b}}(\sqrt{\mid \alpha \mid} t + t_{0}),
\end{equation}
\begin{equation}
\label{eq35}
g = N \exp{\left(\frac{Lbx}{2(b - 2)} + Mx\right)},
\end{equation}
\begin{equation}
\label{eq36}
h = \frac{1}{N}\exp{\left(\frac{Lbx}{2(b - 2)} - Mx\right)},
\end{equation}
where $r = \left(\frac{\ell}{m}\right)^{\frac{(b - 2)}{b}}$. \\
Therefore, we have
\begin{equation}
\label{eq37}
A = E_{0} \exp{\left(\frac{Lx}{(b - 2)}\right)} \cosh^{\frac{(b - 2)}{
2b}}(\sqrt{\mid \alpha \mid} t + t_{0}),
\end{equation}
\begin{equation}
\label{eq38}
B = G_{0} \exp{\left(\frac{Lbx}{2(b - 2)} + Mx\right)} \cosh^{\frac{1}{2}}
(\sqrt{\mid \alpha \mid} t + t_{0}),
\end{equation}
\begin{equation}
\label{eq39}
C = H_{0} \exp{\left(\frac{Lbx}{2(b - 2)} - Mx\right)} \cosh^{\frac{1}{2}}
(\sqrt{\mid \alpha \mid} t + t_{0}),
\end{equation}
where $E_{0} = nr$, $G_{0} = N \ell $, $H_{0} = \frac{\ell}{N}$. \\
After using suitable transformation of coordinates, the metric (\ref{eq1})
reduces to the form
\[
ds^{2} = E_{0}^{2} \exp{\left(\frac{2LX}{(b - 2)}\right)} \cosh^{\frac{(b - 2)}{b}}
(\sqrt{\mid \alpha \mid}
T) (dX^{2} - dT^{2}) +
\]
\[
\exp{\left(\frac{LbX}{(b - 2)} + 2MX \right)} \cosh(\sqrt{\mid \alpha \mid} T)
 dY^{2}+
\]
\begin{equation}
\label{eq40}
\exp{\left(\frac{LbX}{(b - 2)} - 2MX\right)} \cosh(\sqrt{\mid \alpha \mid} T) dZ^{2},
\end{equation}
For the specification of displacement vector $\beta(t)$ within the
framework of Lyra geometry and for realistic models of physical
importance, we consider the following two cases by taking $\beta$ as
constant and also $\beta$ as function of time.
\begin{center}
\large{4.1. \textit{When $\beta$ is a constant i.e. $\beta = \beta_{0}$ (constant)}}
\end{center}
Using Eqs. (\ref{eq37}), (\ref{eq38}) and (\ref{eq39}) in Eqs.
(\ref{eq10}) and (\ref{eq13}) the expressions for pressure $p$ and
density $\rho$ for the model (\ref{eq40}) are given by
\[
8 \pi p = \frac{1}{E_{0}^{2}}\exp{\left(\frac{2LX}{2 - b}\right)}\cosh^{\frac{2 - b}{b}}
{\left(\sqrt{\mid \alpha \mid T}
\right)}\times
\]
\[
\Biggl[\mid \alpha \mid \left\{\frac{(b + 1)}{4b}\tanh^{2}{\left(\sqrt{\mid \alpha
\mid T}\right)} - 1 \right\}
+ \frac{b(b + 4)L^{2}}{4(b - 2)^{2}} - M^{2} + \frac{MLb}{(b - 2)}\Biggr]
\]
\begin{equation}
\label{eq41}
- \frac{3}{4}\beta_{0}^{2},
\end{equation}
\[
8 \pi \rho = \frac{1}{E_{0}^{2}}\exp{\left(\frac{2LX}{2 - b}\right)}
\cosh^{\frac{2 - b}{b}}{\left(\sqrt{\mid \alpha \mid T}
\right)}\times
\]
\begin{equation}
\label{eq42}
\Biggl[\frac{\mid \alpha \mid (3b - 4)}{4b}\tanh^{2}{\left(\sqrt{\mid \alpha
\mid T}\right)}
+ \frac{b(4 - 3b)L^{2}}{4(b - 2)^{2}} - M^{2} + \frac{MLb}{(b - 2)}\Biggr]
+ \frac{3}{4}\beta_{0}^{2}.
\end{equation}
From Eq. (\ref{eq17}) the non-vanishing component $F_{12}$ of the electromagnetic
field tensor is obtained as
\begin{equation}
\label{eq43}
F_{12} = \sqrt{\frac{\bar{\mu}MLb}{4\pi(2 - b)}} ~ G_{0} ~ \exp{\left\{\left(\frac{LB}
{b - 2} + 2M\right)\frac{X}{2}\right\}}
\cosh{\left(\sqrt{\mid \alpha \mid}T\right)}.
\end{equation}
From above equation it is observed that the electromagnetic field tensor increases
with time.\\

The reality conditions (Ellis \cite{ref47})
$$
(i) \rho + p > 0, ~ ~ (ii) \rho + 3p > 0,
$$
lead to
\begin{equation}
\label{eq44}
\frac{(4b - 3)\mid \alpha \mid}{4b}\tanh^{2}{\left(\sqrt{\mid \alpha \mid} T\right)}
 > \mid \alpha \mid + 2 M^{2}
- \frac{2MLb}{(b - 2)} + \frac{L^{2}b(b - 4)}{2(b - 2)^{2}},
\end{equation}
and
\[
\exp{\left(\frac{2LX}{2 - b}\right)}\Biggl[\frac{(6b - 1)\mid \alpha \mid}{4b}
\tanh^{2}{\left(\sqrt{\mid \alpha \mid}T\right)}
- 3 \mid \alpha \mid + \frac{4bL^{2}}{(b - 2)^{2}} - 4M^{2}
\]
\begin{equation}
\label{eq45}
 + \frac{4MLb}{(b - 2)}\Biggr]  > \frac{3}{2}\beta_{0}^{2}\cosh^{\frac{b - 2}{b}}
 {\left(\sqrt{\mid \alpha \mid} T \right)}.
\end{equation}
The dominant energy conditions (Hawking and Ellis \cite{ref48})
$$
(i) \rho - p \geq 0, ~ ~ (ii) \rho + p \geq 0,
$$
lead to
\[
\exp{\left(\frac{2LX}{2 - b}\right)}\Biggl[\frac{(2b - 5)\mid \alpha \mid}{4b}
\tanh^{2}{\left(\sqrt{\mid \alpha \mid} T\right)}
+ \mid \alpha \mid \Biggr]
\]
\begin{equation}
\label{eq46}
 + \frac{3}{2}\beta_{0}^{2}E_{0}^{2}\cosh^{\frac{b - 2}{2}}
 {\left(\sqrt{\mid \alpha \mid} T\right)} \geq
\exp{\left(\frac{2LX}{2 - b}\right)} \frac{L^{2}b^{2}}{(b - 2)^{2}},
\end{equation}
and
\begin{equation}
\label{eq47}
\frac{(4b - 3)\mid \alpha \mid}{4b}\tanh^{2}{\left(\sqrt{\mid \alpha
\mid} T\right)} \geq \mid \alpha \mid + 2 M^{2}
- \frac{2MLb}{(b - 2)} + \frac{L^{2}b(b - 4)}{2(b - 2)^{2}}.
\end{equation}
The conditions (\ref{eq45}) and (\ref{eq46}) impose the restriction
on $\beta_{0}$.
\begin{center}
\large{4.2. \textit{When $\beta$ is a function of $t$}}
\end{center}
In this case to find the explicit value of displacement field
$\beta(t)$, we assume that the fluid obeys an equation of state of
the form
\begin{equation}
\label{eq48}
p = \gamma \rho,
\end{equation}
where $\gamma(0 \leq \gamma \leq 1)$ is a constant. Here we consider three
cases of physical interest.
\begin{center}
\large{4.2.1 \textit{Empty universe}}
\end{center}
Let us consider $\gamma = 0$. In this case $p = \rho = 0$. Thus, from Eqs.
(\ref{eq11}) and (\ref{eq14}) we obtain
\[
\beta^{2}(t) = \frac{2}{3E^{2}}\exp{\left(\frac{2LX}{2 - b}\right)}
\cosh^{\frac{2 - b}{b}}{\left(\sqrt{\mid \alpha \mid}
T\right)}\times
\]
\begin{equation}
\label{eq49}
\left[\frac{L^{2}b^{2}}{(b - 2)^{2}} - \frac{\mid \alpha \mid}{2}
\tanh^{2}{\left(\sqrt{\mid \alpha \mid} T\right)}
- \mid \alpha \mid \right]
\end{equation}

Halford \cite{ref6} has pointed out that the constant vector
displacement field $\phi_i$ in Lyra's geometry plays the role of
cosmological constant $\Lambda$ in the normal general relativistic
treatment. From Eq. (\ref{eq49}), it is observed that the
displacement vector $\beta(t)$ is a decreasing function of time.
\begin{center}
\large{4.2.2. \textit{Zeldovich universe}}
\end{center}
Let us consider $\gamma = 1$. In this case
\begin{equation}
\label{eq50}
\rho = p
\end{equation}
Using Eqs. (\ref{eq37}), (\ref{eq38}) and (\ref{eq39}) in Eqs. (\ref{eq11})
and (\ref{eq14}), we obtain
\[
\beta^{2}(t) = \frac{2}{3E^{2}}\exp{\left(\frac{2LX}{2 - b}\right)}
\cosh^{\frac{2 - b}{b}}{\left(\sqrt{\mid \alpha \mid}
T \right)}\times
\]
\begin{equation}
\label{eq51}
\left[\frac{L^{2}b^{2}}{(b - 2)^{2}} - \frac{\mid \alpha \mid}{2}
\tanh^{2}{\left(\sqrt{\mid \alpha \mid} T\right)}
- \mid \alpha \mid \right]
\end{equation}
From Eq. (\ref{eq51}), it is observed that displacement vector $\beta$ is
decreasing function of time.
The expressions for pressure $p$ and energy density $\rho$ are given by
\[
8\pi p = 8\pi \rho = \frac{1}{E^{2}}\exp{\left(\frac{2LX}{2 - b}\right)}
\cosh^{\frac{2 - b}{b}}{\left(\sqrt{\mid \alpha \mid}
T \right)}\times
\]
\begin{equation}
\label{eq52}
\Biggl[\frac{(2b + 1)\mid \alpha \mid}{b}\tanh^{2}{\left(\sqrt{\mid \alpha
\mid} T\right)} + \frac{b(4 -b)L^{2}}{4(b - 2)^{2}}
+ \frac{MLb}{(b - 2)} - M^{2} - \frac{3\mid \alpha \mid}{2}\Biggr]
\end{equation}
The reality condition (Ellis 1973)
$$
(i) \rho + p > 0, ~ ~ (ii) \rho + 3p > 0,
$$
lead to
\begin{equation}
\label{eq53}
\frac{(2b + 1)\mid \alpha \mid}{b}\tanh^{2}{\left(\sqrt{\mid \alpha \mid}
T\right)} > \frac{b(4 -b)L^{2}}{4(b - 2)^{2}}
- \frac{MLb}{(b - 2)} + M^{2} + \frac{3\mid \alpha \mid}{2}
\end{equation}
\begin{center}
\large{4.2.3. \textit{Radiating universe}}
\end{center}
Let us consider $\gamma = \frac{1}{3}$. In this case
\begin{equation}
\label{eq54}
\rho = 3 p
\end{equation}
In this case using Eqs. (\ref{eq37}) - (\ref{eq39}) in Eqs. (\ref{eq11}) -
(\ref{eq14}), the expressions for
$\beta(t)$, $p$ and $\rho$ are obtained as
\[
\beta^{2}(t) = \frac{2}{3E^{2}}\exp{\left(\frac{2LX}{2 - b}\right)}
\cosh^{\frac{2 - b}{b}}{\left(\sqrt{\mid \alpha \mid}
T \right)}\times
\]
\begin{equation}
\label{eq55}
\left[\frac{L^{2}b^{2}}{2(b - 2)^{2}}  + 2M^{2} - \frac{(b + 4)\mid \alpha
\mid}{2}\tanh^{2}{\left(\sqrt{\mid \alpha \mid}
T\right)} - \frac{(b - 2)\mid \alpha \mid }{2b}\right]
\end{equation}
\[
8\pi p = \frac{1}{E^{2}}\exp{\left(\frac{2LX}{2 - b}\right)}
\cosh^{\frac{2 - b}{b}}{\left(\sqrt{\mid \alpha \mid} T \right)}
\Biggl[\frac{(2b + 5)\mid \alpha \mid}{4b} \tanh^{2}{\sqrt{\mid \alpha \mid} T}
\]
\begin{equation}
\label{eq56}
 + \frac{bL^{2}}{(b - 2)} + \frac{MLb}{4(b - 2)} + \frac{(b - 2)\mid
 \alpha \mid }{2b} - 2M^{2} - \mid \alpha \mid \Biggr]
\end{equation}
\[
8\pi \rho = \frac{3}{E^{2}}\exp{\left(\frac{2LX}{2 - b}\right)}
\cosh^{\frac{2 - b}{b}}{\left(\sqrt{\mid \alpha \mid} T \right)}
\Biggl[\frac{(2b + 5)\mid \alpha \mid}{4b} \tanh^{2}{\sqrt{\mid \alpha \mid} T}
\]
\begin{equation}
\label{eq57}
 + \frac{bL^{2}}{(b - 2)} + \frac{MLb}{4(b - 2)} + \frac{(b - 2)\mid
 \alpha \mid }{2b} - 2M^{2} - \mid \alpha \mid \Biggr]
\end{equation}
From Eq. (\ref{eq55}), it is observed that displacement vector
$\beta$ is decreasing function of time. The reality conditions
(Ellis \cite{ref47})
$$
(i) \rho + p > 0, ~ ~ (ii) \rho + 3p > 0,
$$
and the dominant energy conditions (Hawking and Ellis \cite{ref48})
$$
(i) \rho - p \geq 0, ~ ~ (ii) \rho + p \geq 0,
$$
lead to
$$
\frac{(2b + 5)\mid \alpha \mid}{4b} \tanh^{2}{\sqrt{\mid \alpha \mid} T} >
\frac{bL^{2}}{(2 - b)} + \frac{MLb}{(2 - b)}
+ \frac{(2 - b)\mid \alpha \mid}{2b} + 2M^{2} + \mid \alpha \mid
$$
and
\begin{equation}
\label{eq58}
\frac{(2b + 5)\mid \alpha \mid}{4b} \tanh^{2}{\sqrt{\mid \alpha \mid} T}
\geq \frac{bL^{2}}{(2 - b)} + \frac{MLb}{(2 - b)}
+ \frac{(2 - b)\mid \alpha \mid}{2b} + 2M^{2} + \mid \alpha \mid
\end{equation}
respectively.\\
The expressions for the expansion $\theta$, shear scalar $\sigma^{2}$,
deceleration parameter $q$ and proper
volume V  for the model (\ref{eq40}) are given by
\begin{equation}
\label{eq59}
\theta = \frac{(3b - 2)\sqrt{\mid \alpha \mid}}{2bE}\exp{\left(\frac{LX}{2 - b}
\right)}\cosh^{\frac{2 - b}{2b}}
{\left(\sqrt{\mid \alpha \mid} T \right)}\tanh^{2}{(\sqrt{\mid \alpha \mid} T)},
\end{equation}
\begin{equation}
\label{eq60}
\sigma^{2} = \frac{\mid \alpha \mid}{3b^{2}E^{2}}\exp{\left(\frac{2LX}{2 - b}
\right)}\cosh^{\frac{2 - b}{b}}
{\left(\sqrt{\mid \alpha \mid} T \right)}\tanh^{2}{(\sqrt{\mid \alpha \mid} T)},
\end{equation}
\begin{equation}
\label{eq61}
q =  - \frac{8b(b-1)E^2\left(1+\frac{b-2}{b}\tanh^2(\sqrt{\mid \alpha \mid} T)\right)}
{9(3b-2)^2\exp\left(\frac{2LX}{2-b}\right)\cosh^{2}{\left(\sqrt{\mid \alpha \mid} 
T\right)}\tanh^4{\left(\sqrt{\mid \alpha \mid}T\right)}},
\end{equation}

\begin{figure}
\begin{center}
\includegraphics[width=4.0in]{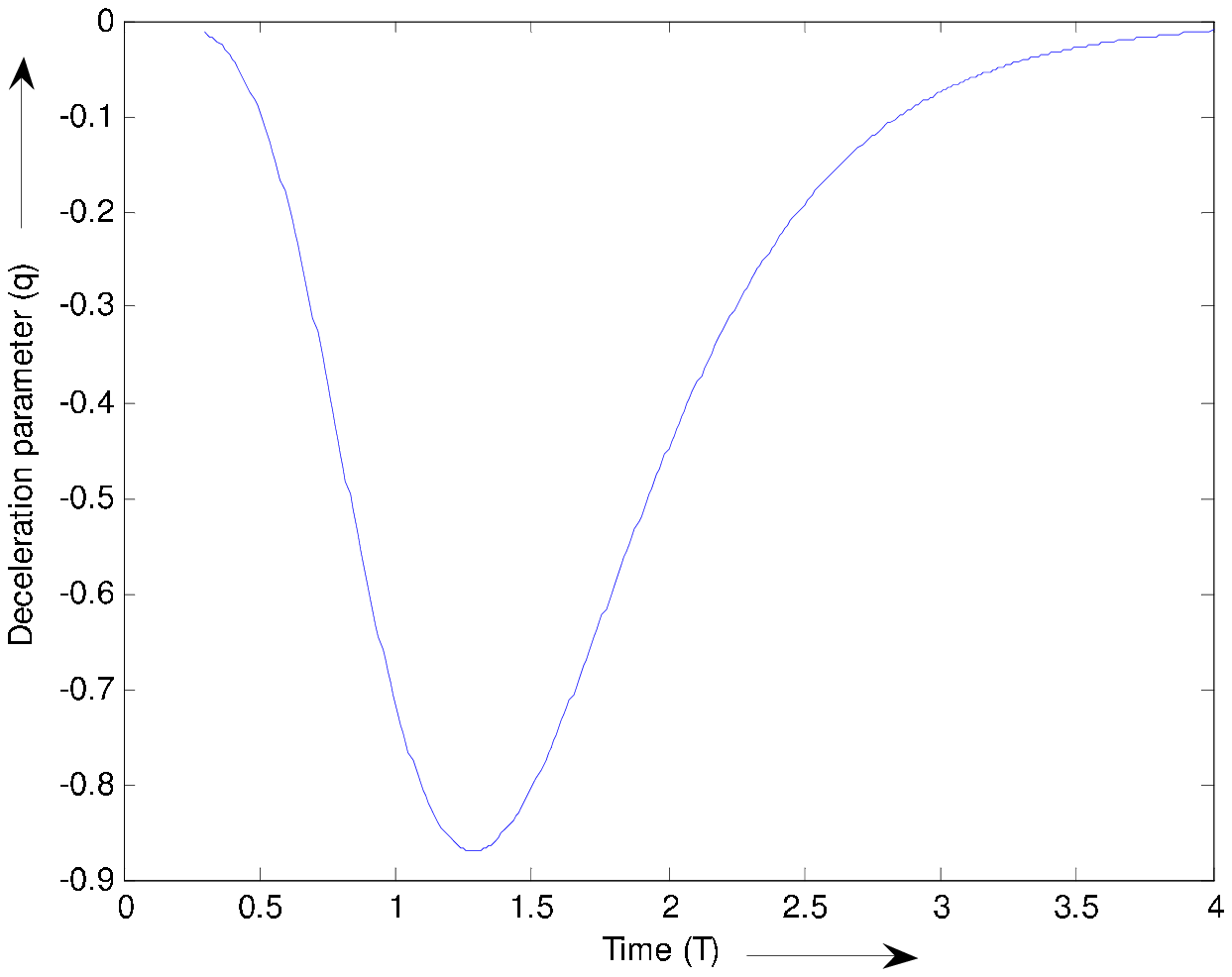} 
\caption{The plot of deceleration parameter (q) vs. time (T).}
\label{fg:Fig1.eps}
\end{center}
\end{figure}

\begin{equation}
\label{eq62}
V = \sqrt{-g} = E^{2} \exp{\left(\frac{(b + 2)LX}{(b - 2)}\right)}
\cosh^{\frac{2(b - 1)}{b}}
{\left(\sqrt{\mid \alpha \mid} T \right)}.
\end{equation}
From Eqs. (\ref{eq59}) and (\ref{eq60}) we obtain
\begin{equation}
\label{eq63}
\frac{\sigma^{2}}{\theta^{2}} = \frac{4}{3(3b - 2)^{2}} =  \mbox{constant}.
\end{equation}
The Hubble parameter $H$  is given by
\begin{equation}
\label{eq64}
H = \frac{3(3b - 2)\sqrt{\mid \alpha \mid}}{2bE}\exp{\left(\frac{LX}{2 - b}
\right)}\cosh^{\frac{2 - b}{2b}}
{\left(\sqrt{\mid \alpha \mid} T \right)}\tanh^{2}{(\sqrt{\mid \alpha \mid} T)},
\end{equation}
The sign of q indicates whether the model inflates or not. A positive sign of q corressponds to standard decelerating
model where as the negative sign $ -1\leq q < 0 $, indicates inflations. Recent observation shows that the 
deceleration parameter of universe is in the range $ -1\leq q < 0 $ and the present universe is undergoing an 
accelerated expansion \cite{ref49, ref50}. This behaviour is clearly shown in Fig. 1, as a representative case with 
appropriate choice of constants and other physical parameters using reasonably well known situation.
Also the current observation of SNe Ia and CMBR favour an accerating model 
$ (q < 0) $. 
From Eq. (\ref{eq61}), it can be seen that the deceleration parameter $q <
0$ when
$$
T < \frac{1}{\sqrt{\mid \alpha \mid}}\tanh^{-1}{\sqrt{\frac{b}{(2 - b)}}}
$$
It follows that our models of the universe are consistent with recent observations.
Generally model (40) represents expanding, shearing and anisotropic universe
in which the flow vector is geodetic.
\begin{center}
\large{4.3. \textit{Thermodynamical behaviour and entropy of universe}}
\end{center}
From the thermodynamics \cite{ref51, ref52}, we apply the combination of first and second law of thermodynamics
to the system with volume V. As we know that
\begin{equation}
 \label{eq65}
 \c{T} dS=d(\rho V)+pdV
\end{equation}
where $ \c{T} $, S represents the tempreture and entropy respectively. \\
Eq. (\ref{eq65}) may be written as
\begin{equation}
 \label{eq66}
 \c{T} dS=d{\left[(\rho + p)V\right]} - Vdp
\end{equation}
The integrability condition is necessary to define a perefect fluid as a thermodynamical 
syetem \cite{ref52}$-$ \cite{ref55}. It is given by
\begin{equation}
 \label{eq67}
 dp=\frac{\rho + P}{\c{T}}d\c{T}
\end{equation}
Plugging eq. (\ref{eq67}) in eq. (\ref{eq66}), we have the differential equation
\begin{equation}
 \label{eq68}
 dS=\frac{1}{\c{T}}d{\left[(\rho + p)V\right]} - (\rho + p)V\frac{d\c{T}}{\c{T}^2}
\end{equation}
we rewrite eq (\ref{eq68}) as
\begin{equation}
 \label{eq69}
 dS=d{\left[\frac{(\rho + p)V}{\c{T}} + c\right]}
\end{equation}
where c is constant.\\
Hence the entropy is defined as
\begin{equation}
 \label{70}
 S=\frac{\rho + p}{\c{T}} V
\end{equation}
Let the entropy density be s, so that
\begin{equation}
 \label{eq71}
 s=\frac{S}{V}=\frac{\rho + p}{\c{T}}=\frac{(1+\gamma)\rho}{\c{T}}
\end{equation}
where $ p=\gamma\rho $ and $ 0<\gamma\leq1 $.\\
If we define the entropy density in terms of temprature then the first law of thermodynamics may be written as
\begin{equation}
 \label{eq72}
 d(\rho V)+\gamma\rho dV=(1+\rho)\c{T}d\left(\frac{\rho V}{\c{T}}\right)
\end{equation}
which on integration yields
\begin{equation}
 \label{73}
 \c{T}= c_{0}\rho^{\frac{\gamma}{(1+\gamma)}}
\end{equation}
where $ c_{0} $ is constant of integration.\\

From eqs. (\ref{eq71}) and (73), we obtain
\begin{equation}
 \label{eq74}
 s=\left(\frac{1+\gamma}{c_{0}}\right)\rho^{\frac{1}{1+\rho}}
\end{equation}
These equation are not valid for $ \gamma=-1 $. For the Zel'dovich fluid $ (\gamma=1) $, we get
\begin{equation}
 \label{eq75}
 \c{T}=c_{0}\rho^\frac{1}{2}
\end{equation}
\begin{equation}
 \label{eq76}
 s=\frac{2}{c_{0}}\rho^\frac{1}{2}
\end{equation}
 $$ \Rightarrow s\sim \rho^\frac{1}{2}\sim \c{T} $$ 
Thus the entropy density is proportional to the tempreture. we have
\[
\c{T} = T_{0}\exp{\left(\frac{LX}{2 - b}\right)}
\cosh^{\frac{2 - b}{2b}}{\left(\sqrt{\mid \alpha \mid}
T \right)}\times
\]
\begin{equation}
\label{eq77}
\Biggl[\frac{(2b + 1)\mid \alpha \mid}{b}\tanh^{2}{\left(\sqrt{\mid \alpha
\mid} T\right)} + \frac{b(4 -b)L^{2}}{4(b - 2)^{2}}
+ \frac{MLb}{(b - 2)} - M^{2} - \frac{3\mid \alpha \mid}{2}\Biggr]^\frac{1}{2}
\end{equation}
\[
s = s_{0}\exp{\left(\frac{LX}{2 - b}\right)}
\cosh^{\frac{2 - b}{2b}}{\left(\sqrt{\mid \alpha \mid}
T \right)}\times
\]
\begin{equation}
\label{eq78}
\Biggl[\frac{(2b + 1)\mid \alpha \mid}{b}\tanh^{2}{\left(\sqrt{\mid \alpha
\mid} T\right)} + \frac{b(4 -b)L^{2}}{4(b - 2)^{2}}
+ \frac{MLb}{(b - 2)} - M^{2} - \frac{3\mid \alpha \mid}{2}\Biggr]^\frac{1}{2}
\end{equation}
\[
S = S_{0}\exp{\left(\frac{(b+1)LX}{2 - b}\right)}
\cosh^{\frac{3b-2}{2b}}{\left(\sqrt{\mid \alpha \mid}
T \right)}\times
\]
\begin{equation}
\label{eq79}
\Biggl[\frac{(2b + 1)\mid \alpha \mid}{b}\tanh^{2}{\left(\sqrt{\mid \alpha
\mid} T\right)} + \frac{b(4 -b)L^{2}}{4(b - 2)^{2}}
+ \frac{MLb}{(b - 2)} - M^{2} - \frac{3\mid \alpha \mid}{2}\Biggr]^\frac{1}{2}
\end{equation}
where $ T_{0}=\frac{c_{0}}{E} $, $ s_{0}=\frac{2}{c_{0}E} $ and $ S_{0}=\frac{2E}{c_{0}} $ are constant.\\
For radiating fluid $ (\gamma=\frac{1}{3}) $, we get
\begin{equation}
 \label{eq80}
 \c{T}\sim\rho^\frac{1}{4}
\end{equation}
\begin{equation}
 \label{eq81}
 s\sim\rho^\frac{3}{4}\sim\c{T}^3
\end{equation} 
Thus the entropy density is proportional to cube of tempreture.\\
Now the tempreture, entropy density and entropy of radiating universe is given by
\[
\c{T} = T_{00}\exp{\left(\frac{LX}{4 - 2b}\right)}
\cosh^{\frac{2 - b}{4b}}{\left(\sqrt{\mid \alpha \mid}
T \right)}\times
\]
\begin{equation}
\label{eq82}
\Biggl[\frac{(2b + 1)\mid \alpha \mid}{b}\tanh^{2}{\left(\sqrt{\mid \alpha
\mid} T\right)} + \frac{b(4 -b)L^{2}}{4(b - 2)^{2}}
+ \frac{MLb}{(b - 2)} - M^{2} - \frac{3\mid \alpha \mid}{2}\Biggr]^\frac{1}{4}
\end{equation}
\[
s = s_{00}\exp{\left(\frac{3LX}{4 - 2b}\right)}
\cosh^{\frac{6 - b}{4b}}{\left(\sqrt{\mid \alpha \mid}
T \right)}\times
\]
\begin{equation}
\label{eq83}
\Biggl[\frac{(2b + 1)\mid \alpha \mid}{b}\tanh^{2}{\left(\sqrt{\mid \alpha
\mid} T\right)} + \frac{b(4 -b)L^{2}}{4(b - 2)^{2}}
+ \frac{MLb}{(b - 2)} - M^{2} - \frac{3\mid \alpha \mid}{2}\Biggr]^\frac{3}{4}
\end{equation}
\[
S = S_{00}\exp{\left(\frac{(2b+1)LX}{2(b-2)}\right)}
\cosh^{\frac{7b+2}{4b}}{\left(\sqrt{\mid \alpha \mid}
T \right)}\times
\]
\begin{equation}
\label{eq84}
\Biggl[\frac{(2b + 1)\mid \alpha \mid}{b}\tanh^{2}{\left(\sqrt{\mid \alpha
\mid} T\right)} + \frac{b(4 -b)L^{2}}{4(b - 2)^{2}}
+ \frac{MLb}{(b - 2)} - M^{2} - \frac{3\mid \alpha \mid}{2}\Biggr]^\frac{3}{4}
\end{equation}
where $ T_{00}=\frac{c_{0}}{\sqrt{E}} $, $ s_{00}=\frac{2}{c_{0}E\sqrt{E}} $ and $ S_{00}=\frac{2\sqrt{E}}{c_{0}} $ 
 are constant.\\
\begin{center}
\large{5. \textit{Case(iii): $ab > 0$}}
\end{center}
In this case we obtain
\begin{equation}
\label{eq85}
f = n \exp{\left(\frac{Lx}{(b - 2)}\right)},
\end{equation}
\begin{equation}
\label{eq86}
\mu = \beta \cos^{\frac{1}{2}}(\sqrt{\alpha} t + t_{0}),
\end{equation}
\begin{equation}
\label{eq87}
\nu = r \cos^{\frac{(b - 2)}{2b}}(\sqrt{\alpha } t + t_{0}),
\end{equation}
\begin{equation}
\label{eq88}
g = N \exp{\left(\frac{Lbx}{2(b - 2)} + Mx\right)},
\end{equation}
\begin{equation}
\label{eq89}
h = \frac{1}{N}\exp{\left(\frac{Lbx}{2(b - 2)} - Mx\right)}.
\end{equation}
Therefore, we have
\begin{equation}
\label{eq90}
A = E \exp{\left(\frac{Lx}{(b - 2)}\right)} \cos^{\frac{(b - 2)}{
2b}}(\sqrt{\alpha} t + t_{0}),
\end{equation}
\begin{equation}
\label{eq91}
B = G \exp{\left(\frac{Lbx}{2(b - 2)} + Mx\right)} \cos^{\frac{1}{2}}
(\sqrt{\alpha } t + t_{0}),
\end{equation}
\begin{equation}
\label{eq92}
C = H \exp{\left(\frac{Lbx}{2(b - 2)} - Mx\right)} \cos^{\frac{1}{2}}
(\sqrt{\alpha} t + t_{0}).
\end{equation}
Here $E$, $G$ and $H$ are already defined in Section 4. \\
After using suitable transformation of coordinates, the metric (\ref{eq1})
reduces to the form
\[
ds^{2} = E^{2} \exp{\left(\frac{2LX}{(b - 2)}\right)} \cos^{\frac{(b - 2)}{b}}
(\sqrt{\alpha}
T) (dX^{2} - dT^{2}) +
\]
\[
\exp{\left(\frac{LbX}{(b - 2)} + 2MX \right)} \cos(\sqrt{\alpha} T) dY^{2}+
\]
\begin{equation}
\label{eq93}
\exp{\left(\frac{LbX}{(b - 2)} - 2MX\right)} \cos(\sqrt{\alpha} T) dZ^{2},
\end{equation}
For the specification of displacement vector $\beta(t)$ within the
framework of Lyra geometry and for realistic models of physical
importance, we consider the following two cases by taking $\beta$ as
constant and also $\beta$ as function of time.
\begin{center}
\large{5.1. \textit{When $\beta$ is a constant i.e. $\beta = \beta_{0}$ (constant)}}
\end{center}
Using Eqs. (\ref{eq90}), (\ref{eq91}) and (\ref{eq92}) in Eqs.
(\ref{eq11}) and (\ref{eq14}) the expressions for pressure $p$ and
density $\rho$ for the model (\ref{eq93}) are given by
\[
8 \pi p = \frac{1}{E^{2}}\exp{\left(\frac{2LX}{2 - b}\right)}\cos^{\frac{2 - b}{b}}
{(\sqrt{\alpha}T)}\times
\]
\begin{equation}
\label{eq94} \Biggl[\alpha \left\{\frac{(3b -
4)}{4b}\tan^{2}{\left(\sqrt{\alpha}T\right)} + 1 \right\} +
\frac{b(b + 4)L^{2}}{4(b - 2)^{2}} - M^{2} + \frac{MLb}{b -
2}\Biggr] - \frac{3}{4}\beta_{0}^{2},
\end{equation}
\[
8 \pi \rho = \frac{1}{E^{2}}\exp{\left(\frac{2LX}{2 - b}\right)}\cos^{\frac{2 - b}{b}}
{(\sqrt{\alpha}T)}\times
\]
\begin{equation}
\label{eq95} 
\Biggl[\frac{\alpha (3b - 4)}{4b}\tan^{2}{\left(\sqrt{\alpha}T\right)} + \frac{b(4 - 3b)
L^{2}}{4(b - 2)^{2}} - M^{2} + \frac{MLb}{b - 2}\Biggr] +
\frac{3}{4}\beta_{0}^{2}.
\end{equation}
From Eq. (\ref{eq17}) the non-vanishing component $F_{12}$ of the electromagnetic
field tensor is obtained as
\begin{equation}
\label{eq96}
F_{12} =  \sqrt{\frac{\bar{\mu}}{8\pi}\frac{2ML}{(2 - b)}} ~ G ~
\exp{\left\{\left(\frac{Lb}{b - 2} + 2M\right)
\frac{X}{2}\right\}}\cos(\sqrt{\alpha}T).
\end{equation}
From above equation it is observed that the electromagnetic field tensor increases
with time.\\

The reality conditions (Ellis \cite{ref47})
$$
(i) \rho + p > 0, ~ ~ (ii) \rho + 3p > 0,
$$
lead to

\begin{equation}
\label{eq97}
\frac{\alpha(3b - 4)}{2b}\tan^{2}{(\sqrt{\alpha}T)} > \frac{b(b - 4)
L^{2}}{2(b - 2)^{2}} + \frac{2MLb}{2 - b}
+ 2M^{2} - \alpha,
\end{equation}

and
\[
\exp{\left(\frac{2LX}{2 - b}\right)}\Biggl[\frac{\alpha(3b - 4)}{b}
\tan^{2}(\sqrt{\alpha}T) + 3\alpha
+ \frac{2bL^{2}}{(b - 2)^{2}} - 4M^{2} + \frac{4MLb}{(b - 2)}\Biggr]
\]
\begin{equation}
\label{eq98}
 > \frac{3}{2}\beta_{0}^{2}E^{2}\cos^{b - 2}{b}{(\sqrt{\alpha}T)}.
\end{equation}
The dominant energy conditions (Hawking and Ellis \cite{ref48})
$$
(i) \rho - p \geq 0, ~ ~ (ii) \rho + p \geq 0,
$$
lead to
\begin{equation}
\label{eq99}
\frac{3}{2}\beta^{2}E^{2}\cos^{\frac{b - 2}{b}}{\left(\sqrt{\alpha}T\right)}
\geq \exp{\left(\frac{2LX}{2 - b}\right)}
\Biggl[\frac{b^{2}L^{2}}{(b - 2)^{2}} + \alpha \Biggr]
\end{equation}
and
\begin{equation}
\label{eq100}
\frac{\alpha(3b - 4)}{2b}\tan^{2}{(\sqrt{\alpha}T)} \geq \frac{b(b - 4)
L^{2}}{2(b - 2)^{2}} + \frac{2MLb}{2 - b}
+ 2M^{2} - \alpha.
\end{equation}
The conditions (\ref{eq108}) and (\ref{eq109}) impose the restriction
on $\beta_{0}$.
\begin{center}
\large{5.2. \textit{When $\beta$ is a function of $t$}}
\end{center}
In this case to find the explicit value of displacement field
$\beta(t)$, we assume that the fluid obeys an equation of state
given by eqs. (\ref{eq48}). Here we consider three cases of physical
interest.
\begin{center}
\large{5.2.1. \textit{Empty universe}}
\end{center}
Let us consider $\gamma = 0$ in (\ref{eq48}). In this case $p = \rho = 0$.
Thus, from Eqs. (\ref{eq11}) and (\ref{eq14}) we obtain
\begin{equation}
\label{eq101}
\beta^{2}(t) = \frac{2}{3E^{2}}\exp{\left(\frac{2LX}{2 - b}\right)}
\cos^{\frac{2 - b}{b}}{(\sqrt{\alpha}T)}
\Biggl[M^{2} + \frac{L^{2}b^{2}}{(b - 2)^{2}} + \alpha \Biggr].
\end{equation}
Halford \cite{ref6} has pointed out that the constant vector
displacement field $\phi_i$ in Lyra's geometry plays the role of
cosmological constant $\Lambda$ in the normal general relativistic
treatment. From Eq. (\ref{eq101}), it is observed that the
displacement vector $\beta(t)$ is a decreasing function of time.
\begin{center}
\large{5.2.2. \textit{Zeldovich Universe}}
\end{center}
Let us consider $\gamma = 1$. Hence, Eq. (\ref{eq48}) gives $\rho = p$.

Using Eqs. (\ref{eq90}), (\ref{eq91}) and (\ref{eq92}) in Eqs. (\ref{eq11})
and (\ref{eq14}), we obtain
\begin{equation}
\label{eq102}
\beta^{2}(t) = \frac{2}{3E^{2}}\exp{\left(\frac{2LX}{2 - b}\right)}
\cos^{\frac{2 - b}{b}}{(\sqrt{\alpha}T)}
\Biggl[M^{2} + \frac{L^{2}b^{2}}{(b - 2)^{2}} + \alpha \Biggr].
\end{equation}
From Eq. (\ref{eq102}), it is observed that displacement vector
$\beta$ is decreasing function of time. The expressions for pressure
$p$ and energy density $\rho$ are given by
\[
8\pi p = 8\pi \rho = \frac{1}{E^{2}}\exp{\left(\frac{2LX}{2 - b}\right)}
\cos^{\frac{2 - b}{b}}{(\sqrt{\alpha}T)}
\Biggl[\frac{(3b - 4)\alpha}{4b}\tan^{2}\left(\sqrt{\alpha}T\right) + \frac{\alpha}{2}
\]
\begin{equation}
\label{eq103}
+ \frac{b(4 - b)L^{2}}{4(b - 2)^{2}} - \frac{3}{2}M^{2} +\frac{MLb}{b - 2}\Biggr]
\end{equation}
The reality condition (Ellis \cite{ref47})
$$
(i) \rho + p > 0, ~ ~ (ii) \rho + 3p > 0,
$$
lead to
\begin{equation}
\label{eq104}
\frac{(3b - 4)\alpha}{4b}\tan^{2}\left(\sqrt{\alpha}T\right) > \frac{b(b - 4)
L^{2}}{4(b - 2)^{2}} + \frac{3}{2}M^{2}
+\frac{MLb}{2 - b} - \frac{\alpha}{2}.
\end{equation}
\begin{center}
\large{5.2.3. \textit{Radiating Universe}}
\end{center}
Let us consider $\gamma = \frac{1}{3}$. Hence, Eq. (\ref{eq48}) reduces to
$\rho = 3 p$.

In this case using Eqs. (\ref{eq90}) - (\ref{eq92}) in Eqs. (\ref{eq11}) -
(\ref{eq14}), the expressions for
$\beta(t)$, $p$ and $\rho$ are obtained as
\[
\beta^{2}(t) = \frac{2}{3E^{2}}\exp{\left(\frac{2LX}{2 - b}\right)}
\cos^{\frac{2 - b}{b}}{(\sqrt{\alpha}T)}
\Biggl[\frac{\alpha(3b - 2)}{2b} +
\]
\begin{equation}
\label{eq105}
\frac{\alpha(5b - 4)}{4b}\tan^{2}\left(\sqrt{\alpha}T\right) + \frac{3b^{2}L^{2}}
{4(b - 2)^{2}} + M^{2}\Biggr]
\end{equation}
\[
8\pi p = \frac{1}{E^{2}}\exp{\left(\frac{2LX}{2 - b}\right)}\cos^{\frac{2 - b}{b}}
{(\sqrt{\alpha}T)}
\Biggl[\frac{MLb}{b - 2} +
\]
\begin{equation}
\label{eq106}
\frac{bL^{2}(4 - 3b)}{4(b - 2)^{2}} - \frac{\alpha}{2b} - 2M^{2} - \frac{1}{2}
\alpha \tan^{2}\left(\sqrt{\alpha} T\right)\Biggr],
\end{equation}
\[
8\pi \rho = \frac{3}{E^{2}}\exp{\left(\frac{2LX}{2 - b}\right)}\cos^{\frac{2 - b}
{b}}{(\sqrt{\alpha}T)}
\Biggl[\frac{MLb}{b - 2} +
\]
\begin{equation}
\label{eq107}
\frac{bL^{2}(4 - 3b)}{4(b - 2)^{2}} - \frac{\alpha}{2b} - 2M^{2} -
\frac{1}{2}\alpha \tan^{2}\left(\sqrt{\alpha} T\right)\Biggr],
\end{equation}
From Eq. (\ref{eq105}), it is observed that displacement vector
$\beta$ is decreasing function of time. The reality conditions
(Ellis \cite{ref47})
$$
(i) \rho + p > 0, ~ ~ (ii) \rho + 3p > 0,
$$
and the dominant energy conditions (Hawking and \cite{ref48})
$$
(i) \rho - p \geq 0, ~ ~ (ii) \rho + p \geq 0,
$$
lead to
$$
\frac{\alpha}{2} \tan^{2}\left(\sqrt{\alpha}T\right) > \frac{MLb}{b - 2} +
\frac{bL^{2}(4 - 3b)}{4(b - 2)^{2}}
- \frac{\alpha}{2b} - 2M^{2}
$$
and
\begin{equation}
\label{eq108}
\frac{\alpha}{2} \tan^{2}\left(\sqrt{\alpha}T\right) \geq \frac{MLb}{b - 2}
+ \frac{bL^{2}(4 - 3b)}{4(b - 2)^{2}}
- \frac{\alpha}{2b} - 2M^{2}
\end{equation}
respectively.\\
The expressions for the expansion $\theta$, shear scalar
$\sigma^{2}$, deceleration parameter $q$ and proper volume V
for the model (\ref{eq103}) are given by
\begin{equation}
\label{eq109}
\theta = \frac{(2 - 3b)\sqrt{\alpha}}{2bE} \exp{\left(\frac{LX}{2 - b}\right)}
\cos^{\frac{2 - b}{2b}}
{\left(\sqrt{\alpha}T\right)}\tan {\left(\sqrt{\alpha}T\right)},
\end{equation}
where $c_{2} = c_{1}^{\frac{(2 - b)}{2b}}$.
\begin{equation}
\label{eq110}
\sigma^{2} = \frac{\alpha}{3b^{2}E^{2}}\exp{\left(\frac{2LX}{2 - b}\right)}
\cos^{\frac{2 - b}{b}}
\left(\sqrt{\alpha}T\right)\tan^{2}{\left(\sqrt{\alpha}T \right)},
\end{equation}
\begin{equation}
\label{eq111}
q = - \frac{8b(b-1)E^2\left(1-\frac{b-2}{b}\tan^2(\sqrt{\mid \alpha \mid} T)\right)}
{9(3b-2)^2\exp\left(\frac{2LX}{2-b}\right)\cos^{2}{\left(\sqrt{\mid \alpha \mid} 
T\right)}\tan^4{\left(\sqrt{\mid \alpha \mid}T\right)}}, 
\end{equation}

\begin{figure}
\begin{center}
\includegraphics[width=4.0in]{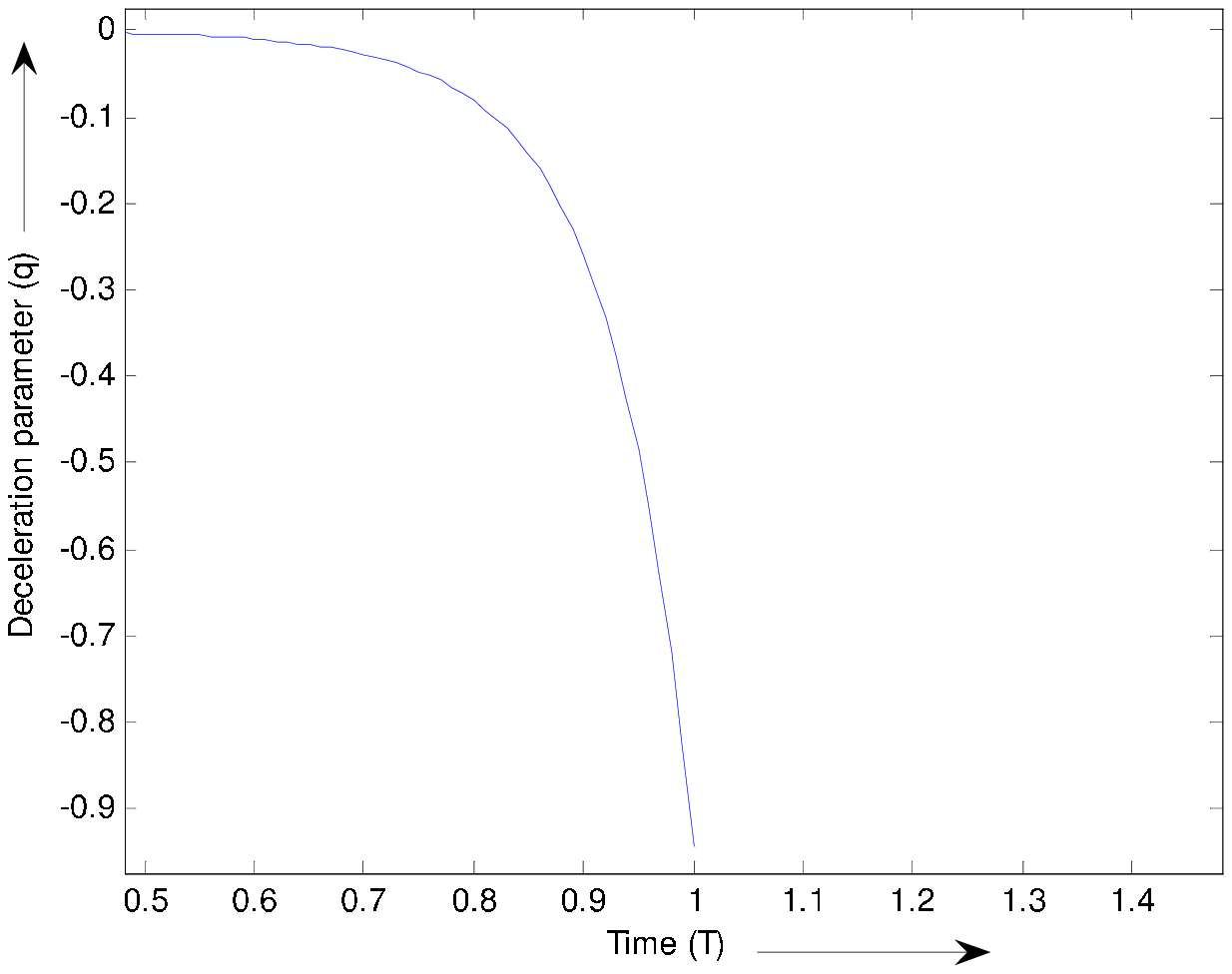} 
\caption{The plot of deceleration parameter (q) vs. time (T).}
\label{fg:Fig2.eps}
\end{center}
\end{figure}

\begin{equation}
\label{eq112}
V = \sqrt{-g} = E^{2} \exp{\left(\frac{(b + 2)LX}{b - 2}\right)}
\cos^{\frac{2(b - 1)}{b}}
\left(\sqrt{\alpha}T\right)
\end{equation}
From Eqs. (\ref{eq109}) and (\ref{eq110}) we obtain
\begin{equation}
\label{eq113}
\frac{\sigma^{2}}{\theta^{2}} = \frac{4}{3(3b - 2)^{2}} =  \mbox{constant}.
\end{equation}
The rotation $\omega$ is identically zero. \\
The Hubble parameter $H$ is given by
\begin{equation}
\label{eq114}
H = \frac{3(2 - 3b)\sqrt{\alpha}}{2bE} \exp{\left(\frac{LX}{2 - b}\right)}
\cos^{\frac{2 - b}{2b}}
{\left(\sqrt{\alpha}T\right)}\tan {\left(\sqrt{\alpha}T\right)},
\end{equation}
For $\frac{2}{b} > 3$, the model (\ref{eq93}) starts expanding at
$T = 0$ and attains its maximum value at $T =
\frac{\pi}{4\sqrt{\alpha}}$. After that $\theta$ decreases to attain
its minimum negative value at $T = \frac{3\pi}{4\sqrt{\alpha}}$.
Thus, the model oscillates with the period
$\frac{\pi}{2\sqrt{\alpha}}$. Since $\frac{\sigma}{\theta} = $
constant, the model does not approach isotropy. The sign of q indicates whether the model inflates or not. 
A positive sign of q corressponds to standard decelerating
model where as the negative sign $ -1\leq q < 0 $, indicates inflations. Recent observation shows that the 
deceleration parameter of universe is in the range $ -1\leq q < 0 $ and the present universe is undergoing an 
accelerated expansion \cite{ref49, ref50}. This behaviour is clearly shown in Fig. 2. Also the current observation of 
SNe Ia and CMBR favour an accerating model $ (q < 0) $.  From Eq.
(\ref{eq111}) it can be seen that the deceleration parameter $q < 0$
when
$$
T < \frac{1}{\sqrt{\alpha}}\tan^{-1}{\left(\sqrt{\frac{b}{(b - 2)}}\right)}.
$$
It follows that our models of the universe are consistent with
recent observations. 
\begin{center}
\large{5.3. \textit{Thermodynamical behaviour and entropy of universe}}
\end{center} 
From eqs.(\ref{eq75}), (76) and (71), the expression for tempreture, entropy density and entropy for zel'dovich fluid
$ (\gamma=1) $ is given by
\[
\c{T} = T_{0}\exp{\left(\frac{LX}{2 - b}\right)}
\cos^{\frac{2 - b}{2b}}{\left(\sqrt{\mid \alpha \mid}
T \right)}
\Biggl[\frac{(3b - 4)\alpha}{4b}\tan^{2}\left(\sqrt{\alpha}T\right) + \frac{\alpha}{2}
\]
\begin{equation}
\label{eq115}
+ \frac{b(4 - b)L^{2}}{4(b - 2)^{2}} - \frac{3}{2}M^{2} +\frac{MLb}{b - 2}\Biggr]^\frac{1}{2}
\end{equation}
\[
s = s_{0}\exp{\left(\frac{LX}{2 - b}\right)}
\cos^{\frac{2 - b}{2b}}{\left(\sqrt{\mid \alpha \mid}
T \right)}
\Biggl[\frac{(3b - 4)\alpha}{4b}\tan^{2}\left(\sqrt{\alpha}T\right) + \frac{\alpha}{2}
\]
\begin{equation}
\label{eq116}
+ \frac{b(4 - b)L^{2}}{4(b - 2)^{2}} - \frac{3}{2}M^{2} +\frac{MLb}{b - 2}\Biggr]^\frac{1}{2}
\end{equation}
\[
S = S_{0}\exp{\left(\frac{(b+1)LX}{2 - b}\right)}
\cos^{\frac{3b-2}{2b}}{\left(\sqrt{\mid \alpha \mid}
T \right)}
\Biggl[\frac{(3b - 4)\alpha}{4b}\tan^{2}\left(\sqrt{\alpha}T\right) + \frac{\alpha}{2}
\]
\begin{equation}
\label{eq117}
+ \frac{b(4 - b)L^{2}}{4(b - 2)^{2}} - \frac{3}{2}M^{2} +\frac{MLb}{b - 2}\Biggr]^\frac{1}{2}
\end{equation}
where $ T_{0} $, $ s_{0} $ and $ S_{0} $ are already defined in section $ 4.3 $.\\
From eqs.(\ref{eq80}), (81) and (71), the expression for tempreture, entropy density and entropy for radiating fluid
$ (\gamma=\frac{1}{3}) $ is given by
\[
\c{T} = T_{00}\exp{\left(\frac{LX}{4 - 2b}\right)}
\cos^{\frac{2 - b}{4b}}{\left(\sqrt{\mid \alpha \mid}
T \right)}
\Biggl[\frac{(3b - 4)\alpha}{4b}\tan^{2}\left(\sqrt{\alpha}T\right) + \frac{\alpha}{2}
\]
\begin{equation}
\label{eq118}
+ \frac{b(4 - b)L^{2}}{4(b - 2)^{2}} - \frac{3}{2}M^{2} +\frac{MLb}{b - 2}\Biggr]^\frac{1}{4}
\end{equation}
\[
s = s_{00}\exp{\left(\frac{3LX}{4 - 2b}\right)}
\cos^{\frac{6 - b}{4b}}{\left(\sqrt{\mid \alpha \mid}
T \right)}
\Biggl[\frac{(3b - 4)\alpha}{4b}\tan^{2}\left(\sqrt{\alpha}T\right) + \frac{\alpha}{2}
\]
\begin{equation}
\label{eq119}
+ \frac{b(4 - b)L^{2}}{4(b - 2)^{2}} - \frac{3}{2}M^{2} +\frac{MLb}{b - 2}\Biggr]^\frac{3}{4}
\end{equation}
\[
S = S_{00}\exp{\left(\frac{(2b+1)LX}{2(b-2)}\right)}
\cos^{\frac{7b+2}{4b}}{\left(\sqrt{\mid \alpha \mid}
T \right)}
\Biggl[\frac{(3b - 4)\alpha}{4b}\tan^{2}\left(\sqrt{\alpha}T\right) + \frac{\alpha}{2}
\]
\begin{equation}
\label{eq120}
+ \frac{b(4 - b)L^{2}}{4(b - 2)^{2}} - \frac{3}{2}M^{2} +\frac{MLb}{b - 2}\Biggr]^\frac{3}{4}
\end{equation}
where $ T_{00} $, $ s_{00} $ and $ S_{00} $ are already defined in section $ 4.3 $. 
\begin{center}
\large{6. \textit{Case(ii): $ab = 0$}}
\end{center}
In this case we obtain
\begin{equation}
\label{eq121}
f = n \exp{\left(\frac{Lx}{(b - 2)}\right)},
\end{equation}
\begin{equation}
\label{eq122}
\mu = (c_{1}t + t_{0})^{\frac{1}{2}},
\end{equation}
\begin{equation}
\label{eq123}
\nu = \left(\frac{1}{m}\right)^{\frac{(b - 2)}{b}}(c_{1}t + t_{0})
^{\frac{(b - 2)}{2b}},
\end{equation}
\begin{equation}
\label{eq124}
g = N \exp{\left(\frac{Lbx}{2(b - 2)} + Mx\right)},
\end{equation}
\begin{equation}
\label{eq125}
h = \frac{1}{N}\exp{\left(\frac{Lbx}{2(b - 2)} - Mx\right)}.
\end{equation}
Therefore, we have
\begin{equation}
\label{eq126}
A = E_{0} \exp{\left(\frac{Lx}{(b - 2)}\right)} (c_{1} t + t_{0})
^{\frac{(b - 2)}{2b}},
\end{equation}
\begin{equation}
\label{eq127}
B = N \exp{\left(\frac{Lbx}{2(b - 2)} + Mx\right)} (c_{1} t + t_{0})
^{\frac{1}{2}},
\end{equation}
\begin{equation}
\label{eq128}
C = H_{0} \exp{\left(\frac{Lbx}{2(b - 2)} - Mx\right)} (c_{1} t + t_{0})
^{\frac{1}{2}},
\end{equation}
where $E_{0} = n m^{\frac{(2 - b)}{b}}$, $H_{0} = \frac{1}{N}$. \\
After using suitable transformation of coordinates, the metric (\ref{eq1})
reduces to the form
\[
ds^{2} = E^{2}_{0} \exp{\left(\frac{2LX}{(b - 2)}\right)} (c_{1} T)
^{\frac{(b - 2)}{b}} (dX^{2} - dT^{2}) +
\]
\begin{equation}
\label{eq129}
\exp{\left(\frac{LbX}{(b - 2)} + 2MX \right)} (c_{1}T) dY^{2} +
\exp{\left(\frac{LbX}{(b - 2)}
- 2MX\right)}(c_{1}T) dZ^{2},
\end{equation}
It is observed that model (129) is same as obtained by Pradhan and Shyam Sunder \cite{ref34}.Thus the physical and
geometrical properties of model are similar to the model obtained by Pradhan and Shyam sundar \cite{ref34}.
\begin{center}
\large{7. \textit{Discussion and concluding remarks}} 
\end{center}
In the present study, I have investigated the plane symmetric inhomogeneous cosmological models of perfect 
fluid distribution with electromagnetic field based on Lyra's geometry.
The source of the magnetic field is
due to an electric current produced along the z-axis. The free gravitational field is assumed
to be of Petrov-type II non degenerate. It is observed that the gauge function $ \beta(t) $ is large
in begining and reduces fast with the evolution of universe for all cases. Also it is found that 
$ \beta(t) $ decreases as time increases therefore $ \beta(t) $ is decreasing function of time and it play the 
same role in Lyra's geometry as as cosmological constant $ \Lambda(t) $ in general relativity. It means that
the displacement vector $ \beta (t) $ coincides with the nature of cosmological constant $ \Lambda $. 
The nontrivial role of vacuum in the early universe generates $ \Lambda $-term that lead to inflationary phase. 
Therefore the study of cosmological modelsin Lyra's geometry may be relavant to inflationary models. 
There seems a good possibility of Lyra's geometry to provide a theorectical foundation of relativistic 
gravitation and cosmology. However, the astrophysical bodies is still
an open question. In fact, it needs a fair trail for experiment.
It is seen that solution obtained by Pradhan, Yadav and Singh \cite{ref34} are particular case of my solution.
Generally models represent expanding, shearing and Petrov type II non degenerate universe in which flow vector
is geodetic. Also $ \frac{\sigma}{\theta}\neq 0 $, thus models do not approach to isotropy. The value of deceleration 
parameter (q) is found to negative $\left(-1 < q < 0\right)$, implying that our universe is accelerating which is 
supported by SNe Ia and CMBR observations. This behavior is clearly depicated in Figure 1 and 2. From Fig. 1, we 
note that at early stage of universe deceleration parameter (q) oscillates between $ q < 0 $ and $ q > -1 $ and 
afterwards it will be uniform (negative) for ever. This has physical meaning.   
\newline
\par
The idea of premordinal magnetism is appealing because it can potentially explain all large scale fields
seen in the universe todays, especially those found in remote proto galaxies. As a result, the literature
contains many studies that examine the role and implications of magnetic field in cosmology.
Maarteens \cite{ref56} in his study explained that magnetic
fields are observed not only in stars but also in galaxies. In
princple, these fields could play a significant role in structure
formation but also affect the anisotropies in cosmic microwave
background radiation [CMB]. Since the electric and magnetic fields
are interrelated, their independent nature disappears when we
consider them as time dependance. Hence, it would be proper to look
upon these fields as a single field - electromagnetic field.
It is worth mentioning here
that magnetic field affects all the physical and kinematical quantities but it does not affect the rate of expansion.
Also we see that in absence of magnetic field, inhomogeneity of universe dies out. This signifies the role of 
magnetic field. The present study also extend the work of Yadav and Bagora \cite{ref57} with in the framework of
Lyra's geometry and clarify thermodynamics of plane symmetric universe by introducing the integrability condition and 
tempreture. A new general equation of state describing the Zel'dovich fluid and radiating fluid models as a function of 
tempreture and volume is found. The basic equations of thermodynamics for plane symmetric universe has been deduced
which may be useful for better understanding of evolution of universe.    
\begin{center}
\textit{Acknowledgements}
\end{center}
\noindent The author would like to thank the Harish-Chandra Research
Institute, Allahabad, India for hospitality where part of this work is carried out.
The author is grateful to the referee 
for his fruitful comments and suggestions for the improvement of the paper.
The author is also thankful to his wife Anju Yadav for her heartiest co-operation and 
supports.

\end{document}